\documentclass{iau} 
\usepackage{graphicx,amssymb}

\newcommand\arcmin{\mbox{$^\prime$}}%
\newcommand\arcsec{\mbox{$^{\prime\prime}$}}%
\newcommand\fs{\mbox{$.\!\!^{\mathrm s}$}}%

\title[X-ray sources in old Galactic star clusters] 
{X-ray sources in Galactic globular clusters and old open clusters}

\author[Maureen van den Berg]   
{Maureen van den Berg}

\affiliation{Harvard-Smithsonian Center for Astrophysics, 60 Garden Street, Cambridge, MA 02138, USA\\email: {\tt mvandenberg@cfa.harvard.edu}}
   
\pubyear{2019}
\volume{351}  
\jname{Star Clusters: From the Milky Way to the Early Universe}
\editors{A. Bragaglia, M.B. Davies, A. Sills \& E. Vesperini, eds.}
\begin{document}

\maketitle

\begin{abstract}
The features and make up of the population of X-ray sources in
Galactic star clusters reflect the properties of the underlying
stellar environment. Cluster age, mass, stellar encounter rate, binary
frequency, metallicity, and maybe other properties as well, determine
to what extent we can expect a contribution to the cluster X-ray
emission from low-mass X-ray binaries, millisecond pulsars,
cataclysmic variables, and magnetically active binaries. Sensitive
X-ray observations with {\em XMM-Newton} and certainly {\em Chandra}
have yielded new insights into the nature of individual sources and
the effects of dynamical encounters. They have also provided a new
perspective on the collective X-ray properties of clusters, in which
the X-ray emissivities of globular clusters and old open clusters can
be compared to each other and to those of other environments. I will
review our current understanding of cluster X-ray sources, focusing on
star clusters older than about 1 Gyr, illustrated with recent results.
\end{abstract}


\firstsection 

\section{Introduction}

X-ray observations of old star clusters efficiently select the
populations of close interacting binaries. X-ray studies can serve as
a tool to study cluster dynamics: the high stellar densities can give
rise to stellar encounters that may destroy or modify existing
binaries, or create new ones that are rare in the field of the
Galaxy. With X-ray studies we can also investigate specific phases of
binary evolution; the known distance and age are helpful for piecing
together a binary's past and predict its future. I review several
recent results on X-ray sources in old ($\gtrsim1$ Gyr) Galactic star
clusters---both globular clusters and old open clusters---and briefly
summarize earlier work to provide context. I cover the following
subjects: observational studies of luminous X-ray sources in globular
clusters ($\gtrsim$$10^{36}$ erg s$^{-1}$; Section \ref{sec_lum}) and
of sources fainter than $\sim$$10^{33}$ erg s$^{-1}$ (Section
\ref{sec_faint}), and a comparison of the X-ray emissivity of old
stellar populations (Section \ref{sec_xem}). \cite[Verbunt \& Lewin
  (2006)]{verblew06} give a comprehensive overview of globular-cluster
X-ray sources, and more recent updates can be found in, e.g.,
\cite[Heinke (2010)]{hein11}, \cite[Pooley (2016)]{pool16}, and
\cite[Ivanova et al.~(2017)]{ivan17}. \cite[Van den Berg
  (2013)]{vdberg13} gives an overview of X-ray sources in old open
clusters.

\section{Luminous X-ray sources and transients} \label{sec_lum}

The first X-ray sources in globular clusters were discovered in the
early seventies, when the {\em UHURU} and OSO-7 satellites made the
first systematic scans of the X-ray sky (e.g.~\cite[Giacconi et
  al.~1974]{giacea74}, \cite[Clark et al.~1975]{clarea75}). Based on
their X-ray luminosities of $\sim$10$^{36}$ erg s$^{-1}$ or higher,
and their X-ray variability, these cluster sources were thought to be
compact objects---neutron stars or black holes---accreting from a
companion in low-mass X-ray binaries (LMXBs). Such systems had already
been found in other parts of the Galaxy (see \cite[Blumenthal \&
  Tucker (1974)]{blumtuck74} for an early review). The brief bursts of
X-rays (\cite[Grindlay \& Heise 1975]{grinhei75}, \cite[Clark et
  al.~1976]{clarea76}), sometimes detected on top of more steady X-ray
emission, revealed that the compact objects had to be neutron stars
(see e.g.~review by \cite[Lewin et al.~1995]{lewiea95}). By
considering the total mass enclosed in the globular clusters of our
Galaxy, it was soon recognized that the number of bright X-ray sources
associated with globular clusters is disproportionately high
(\cite[Katz 1975]{katz75}). Since then, the luminous X-ray sources in
globular clusters have been linked to dynamical interactions (tidal
capture, collisions, exchange encounters; see \cite[Verbunt \& Lewin
  2006]{verblew06} for references) that must be operating uniquely in
dense environments. \cite[Verbunt \& Hut (1987)]{verbhut87} calculated
the encounter number $\Gamma$, or the volume integral (over the
cluster core) of $\rho_c^2 / v$ where $\rho_c$ is the central density
and $v$ is the velocity dispersion. The observations clearly support a
correlation between $\Gamma$ and whether a cluster hosts a bright
X-ray source or not, both under the simplifying assumptions for the
cluster structural parameters made initially, and with a more
sophisticated approach (\cite[Bahramian et
  al.~2013]{bahrea13}). $\Gamma$ has also been found to scale with the
number of radio millisecond pulsars (MSPs; \cite[Bahramian et
  al.~2013]{bahrea13}), which are the descendants of LMXBs, in
contrast to the findings by \cite[Bagchi et al.~(2011)]{bagcea11} who
used a more simple calculation for $\Gamma$.

At the time of writing, we know of 21 luminous ($L_X\gtrsim10^{36}$
erg s$^{-1}$) X-ray sources in fifteen globular clusters. These
include eight persistently bright sources (in seven clusters), and
thirteen transients (in eight clusters) that have been observed to
switch between a faint quiescent and a bright state (or vice versa) at
least once. Table~\ref{tab_bright} gives an overview with various
updates compared to the compilation by \cite[Bahramian et
  al.~(2014)]{bahrea14}. Objects in the growing class of very faint
X-ray transients (VFXTs; not included in the table) reach peak
luminosities of $\sim$10$^{34-36}$ erg s$^{-1}$ (\cite[Wijnands et
  al.~2006]{wijnea06}). It is currently not clear whether the VFXTs,
identified in the Galactic Center region but also in globular clusters
(\cite[Heinke et al.~2009]{heinea09}, \cite[Arnason et
  al.~2015]{arnaea15}) form a homogeneous class, and why their
outbursts only reach the observed levels.

For a long time it was assumed that a single X-ray source is
responsible for the bright X-ray emission from a globular cluster.
The unprecedented spatial resolution of the {\em Chandra X-ray
  Observatory} has shown this assumption to be false. Using {\em
  Chandra}, \cite[White \& Angelini (2001)]{whitange01} resolved the
bright source in M\,15 into two bright sources, thereby solving a
long-standing puzzle regarding the source's seemingly conflicting
characteristics. Rapid-response {\em Chandra} observations to follow
up on triggers from X-ray satellites with lower spatial resolution
have also revealed that globular clusters can host multiple transient
LMXBs, such as in Terzan\,5 and NGC\,6440 (see references in
\cite[Pooley 2016]{pool16}). In some cases, the quiescent counterpart
of a bright transient had already been identified in deep pre-outburst
{\em Chandra} images---but, due to limited X-ray sensitivity, not
always. The latest transient is located in Liller\,1, one of the most
dense and most massive globular clusters in the Galaxy. Liller\,1 was
already known to host at least one bright transient, viz.~the Rapid
Burster. In 2018, an X-ray transient went off in Liller\,1 that did
not show the characteristic behavior of the Rapid Burster. With {\em
  Chandra}, it was found that the position of the new transient
coincided with none of the known sources (\cite[Homan et
  al.~2018]{homaea18}). This is an excellent example of the power of
{\em Chandra}: the two X-ray binaries are separated by only
1.4\arcsec. The recent transient and X-ray millisecond pulsar
IGR\,J18245$-$2452 in M\,28 has been ground breaking in the sense that
its crossidentificaton with the known radio millisecond pulsar M\,28~I
provided the first observational proof for the long-assumed formation
scenario for millisecond pulsars, in which an old neutron star is spun
up by accretion in an LMXB (\cite[Papitto et
  al.~2013]{pappea13}). Another interesting system is the transient
IGR\,J17361$-$4441 in NGC\,6388, with an exceptionally hard X-ray
spectrum for a neutron-star LMXB. \cite[Del Santo et
  al.~(2014)]{delsea14} suggested that the transient originated from
the tidal disruption of a planet by a white dwarf, but an LMXB nature
cannot be excluded (\cite[Wijnands et al.~2015]{wijnea15}).

Small-number statistics and, in some cases, poorly known observational
selection biases (e.g.\,related to ill-constrained outburst duty
cycles, as discussed in e.g.\,\cite[Carbone \& Wijnands
  2019]{carbwijn19}), complicate the comparison of the population of
bright LMXBs in Galactic globular clusters with the LMXB populations
in the rest of the Galaxy or in other galaxies. Such comparisons are
relevant for, e.g., constraining the effects of metallicity on the
LMXB formation efficiency or the frequency of ultra-compact LMXBs
(i.e.\,those with orbital periods $<$1 hour).

\begin{table}
  \caption{Luminous X-ray sources and transients in Galactic globular clusters$^1$}
  \label{tab_bright}
  \scriptsize{
  \begin{tabular}{llllll}
    \hline
    Cluster   & ID                                 & Position RA, Dec (J2000) & $P_b$ & Opt & Note \\
    \hline
    Liller\,1 & Rapid Burster [1]                  & 17$^{\rm h}$33$^{\rm m}$24$\fs$61, $-$33$^\circ$23$\arcmin$19.8$\arcsec$ [2;R]  & $-$ & $-$ & T \\
    Liller\,1 & CXOU\,J173324.6$-$332321 [3]       & 17$^{\rm h}$33$^{\rm m}$24$\fs$61, $-$33$^\circ$23$\arcmin$21.2$\arcsec$ [4;X] & $-$ & $-$ & T \\
    Liller\,1 & CXOU\,J173324.1$-$332316 [5]       & 17$^{\rm h}$33$^{\rm m}$24$\fs$14, $-$33$^\circ$23$\arcmin$16.0$\arcsec$ [5;X] & $-$ & $-$ & T?$^a$\\
    M\,15     & M\,15 X-1, AC\,211$^{b}$ [6]       & 21$^{\rm h}$29$^{\rm m}$58$\fs$31, +12$^{\circ}$10$\arcmin$02.89$\arcsec$ [7;R] & 17.1 h [8] &  + [9] & P\\
    M\,15     & M\,15 X-2 [6]                      & 21$^{\rm h}$29$^{\rm m}$58$\fs$13, +12$^{\circ}$10$\arcmin$02.6$\arcsec$ [6;X] & 22.6 m [10] & + [6] & P\\
    M\,28     & IGR\,J18245$-$2452 [11]            & 18$^{\rm h}$24$^{\rm m}$32$\fs$51, $-$24$^{\circ}$52$\arcmin$07.9$\arcsec$ [12;R] & 11.0 h [13] & + [14] & T$^{c}$\\
    NGC\,1851 & 4U\,0513$-$40 [15]                  & 05$^{\rm h}$14$^{\rm m}$06$\fs$48, $-$40$^{\circ}$02$\arcmin$38.8$\arcsec$ [16;X] & $\sim$17 m [17] & + [18] & P\\
    NGC\,2808 & Swift\,J0911$-$6452 [19]            & 09$^{\rm h}$12$^{\rm m}$02$\fs$46, $-$64$^{\circ}$52$\arcmin$06.4$\arcsec$ [20;X] & 44.3 m [21] & $-$ & T\\
    NGC\,6388 & IGR\,J17361$-$4441 [22]            & 17$^{\rm h}$36$^{\rm m}$17$\fs$42, $-$44$^{\circ}$44$\arcmin$05.98$\arcsec$ [23;X] & $-$ & $-$ & T$^{d}$ \\
    NGC\,6440 & SAX\,J1748.9$-$2021$^{e}$ [24]     & 17$^{\rm h}$48$^{\rm m}$52$\fs$2,  $-$20$^{\circ}$21$\arcmin$32.6$\arcsec$ [25;O] & 8.7 h [26] & $+$ [27,28] & T \\
    NGC\,6440 & NGC\,6440 X-2 [29]                 & 17$^{\rm h}$48$^{\rm m}$52$\fs$76, $-$20$^{\circ}$21$\arcmin$24.0$\arcsec$ [30;X] & 57.3 m [31] & $-$  & T\\
    NGC\,6441 & 4U\,1746$-$37 [32]                 & 17$^{\rm h}$50$^{\rm m}$12$\fs$73, $-$37$^{\circ}$03$\arcmin$06.5$\arcsec$ [33;O] & 5.2 h [34] & + [35,33] & P \\ 
    NGC\,6624 & 4U\,1820$-$30 [32]                 & 18$^{\rm h}$23$^{\rm m}$40$\fs$57, $-$30$^{\circ}$21$\arcmin$39.92$\arcsec$ [36;O] & 11.4 m [37] & + [36] & P \\
    NGC\,6652 & XB\,1832$-$330 [38]                & 18$^{\rm h}$35$^{\rm m}$43$\fs$67, $-$32$^{\circ}$59$\arcmin$26.3$\arcsec$ [39;X] & 2.1 h [40] & + [41] & P$^{f}$\\
    NGC\,6712 & 4U\,1850$-$087 [42]                & 18$^{\rm h}$53$^{\rm m}$04$\fs$91, $-$08$^{\circ}$42$\arcmin$19.35$\arcsec$ [43;O] & 20.6 m [44] & + [45] & P\\
    Terzan\,1 & XB\,1732$-$304 [46]                & 17$^{\rm h}$35$^{\rm m}$47$\fs$26, $-$30$^{\circ}$28$\arcmin$55.3$\arcsec$$^g$ [47;X] & $-$ & $-$ & T \\
    Terzan\,2 & 4U\,1722$-$30 [48]                 & 17$^{\rm h}$27$^{\rm m}$33$\fs$15, $-$30$^{\circ}$48$\arcmin$07.8$\arcsec$ [49;X] & $-$$^h$ & $-$ & P \\
    Terzan\,5 & Terzan\,5 X-1$^i$ [50,51]           & 17$^{\rm h}$48$^{\rm m}$05$\fs$23, $-$24$^{\circ}$46$\arcmin$47.6$\arcsec$ [52;O] & $-$ & + [52] & T \\
    Terzan\,5 & Terzan\,5 X-2$^j$ [53,54]           & 17$^{\rm h}$48$^{\rm h}$04$\fs$82, $-$24$^{\circ}$46$\arcmin$48.9$\arcsec$ [55;M] & 21.3 h [56] & + [57] & T \\
    Terzan\,5 & Terzan\,5 X-3$^k$ [58]    & 17$^{\rm h}$48$^{\rm m}$05$\fs$41, $-$24$^{\circ}$46$\arcmin$38.0$\arcsec$ [59;X] & $-$ & $-$ & T \\
    Terzan\,6 & GRS\,1747$-$312 [60]                & 17$^{\rm h}$50$^{\rm m}$46$\fs$86, $-$31$^\circ$16$\arcmin$28.9$\arcsec$ [61;X] & 12.4 h [61] & $-$ & T \\
  \hline
  \end{tabular}
  }
  \vspace{1mm}

  \scriptsize{$^1$Status as of August 2019. The column ``ID'' gives
    (one of) the common source name(s) and a reference to an early
    report of the detection as a persistent or transient luminous
    source. The column ``Position'' gives a recent accurate position
    reported in the literature and the method to measure it: M=Moon
    occultation, O=optical, R=Radio, X=X-ray. The column ``$P_b$'' gives
    the orbital period, and the entry in the column ``Opt'' indicates
    whether an optical (or near-ultraviolet, or near-infrared)
    counterpart has been identified (+) or not ($-$) . The column
    ``Note'' has additional comments and indicates whether a source
    has been persistently luminous since its discovery (P), or whether
    it is transient (T). $^{a}$The {\em Chandra} position of the faint
    ($L_X$(0.5--2.5 keV)$\approx10^{32-34}$ erg s$^{-1}$) source C1 in
    [5] is a positional match to the {\em Einstein} position reported
    for the X-ray source in Liller\,1 (Grindlay et al.~1984),
    suggesting that C1 could be a third transient the
    cluster. $^{b}$AC\,211 is the name of the optical counterpart to
    4U\,2127+119 [9], which is the bright X-ray source in M\,15 that
    was later resolved into M\,15 X-1 and M\,15 X-2 by
    [6]. $^c$transitional MSP, switches between a rotation-powered
    radio pulsar state to an accretion-powered X-ray pulsar state [13]
    $^d$Del Santo et al.~(2014) suggest the source of the transient
    X-ray emission could be the tidal disruption of a planet by a
    white dwarf, but Wijnands et al.~(2015) argue it is more likely
    that this source is an unusual LMXB. $^e$X-ray emission from
    NGC\,6440 was already detected in the 1970s, but it is unclear
    which X-ray source was responsible for this emission. $^f$Johnston
    et al.~(1996) refer to the source as a ``bright X-ray
    transient''. $^g$This position is for the {\em Chandra} source
    CX\,2, the most likely quiescent counterpart to the luminous
    source that has dimmed significantly since the second half of the
    1990s.  $^h$candidate ultra-compact binary (in 't Zand et
    al.~2007) $^i$An often-used name for this transient is
    EXO\,1745$-$248, which is a little unfortunate since it is unclear
    whether Terzan\,5 X-1 is the same source that was detected by
    EXOSAT. $^j$also known as IGR\,J17480$-$2446 $^k$also known as
    Swift\,J174805.3$-$244637.

    \vspace{0.1cm} {\it References}: {\bf [1]} Lewin et al.~1976, IAUC
    2922; {\bf [2]} Moore et al.~2000, ApJ, 532, 1181; {\bf [3]} Homan
    et al.~2018, ATel 11598; {\bf [4]} Bahramian et al.~2018, ATel
    11646; the {\em Chandra} position from [3] was adjusted so that
    the position for the Rapid Burster from [3] matches the more
    accurate radio position from [2]; {\bf [5]} Homer et al.~2001, AJ,
    122, 2627; {\bf [6]} White \& Angelini 2001, ApJ, 561, L101; {\bf
      [7]} Kulkarni et al.~1990, ApJ, 363, L5; {\bf [8]} Ilovaisky et
    al.~1993, A\&A, 270, 139; {\bf [9]} Auri{\`e}re et al.~1984, A\&A,
    138, 415; {\bf [10]} Dieball et al.~2005, ApJ, 634, L105; {\bf
      [11]} Eckert et al.~2013, ATel 4925; {\bf [12]} Pavan et
    al.~2013, ATel 4981; {\bf [13]} Papitto et al.~2013, Nature, 501,
    517; {\bf [14]} Pallanca et al.~2013, ATel 5003; {\bf [15]} Clark
    et al.~1975, ApJ, 199, L93; {\bf [16]} Fiocchi et al.~2011, MNRAS,
    414, L41; {\bf [17]} Zurek et al.~2009, ApJ, 699, 1113; {\bf [18]}
    Homer et al.~2001, ApJ, 550, L155; {\bf [19]} Serino et al.~2016,
    ATel 8862; {\bf [20]} Homan et al.~2016, ATel 8971; {\bf [21]}
    Sanna et al.~2017, A\&A, 598, A34; {\bf [22]} Gibaud et al.~2011,
    ATel 3565; {\bf [23]} Pooley et al.~2011, ATel 3627; {\bf [24]} in
    't Zand et al.~1999, A\&A, 345, 100; {\bf [25]} Cadelano et
    al.~2017, ApJ, 844, 53; {\bf [26]} Altamirano et al.~2008, ApJ,
    674, L45; {\bf [27]} Verbunt et al.~2000, A\&A, 359, 960; {\bf
      [28]} in 't Zand et al.~2001, ApJ, 563, L41; {\bf [29]} Heinke
    \& Budac 2009, ATel 2139; {\bf [30]} Heinke et al.~2010, ApJ, 714,
    894; {\bf [31]} Altamirano et al.~2010, ApJ, 712, 58; {\bf [32]}
    Giacconi et al.~1974, ApJS 27, 37; {\bf [33]} Homer et al.~2002,
    AJ, 123, 3255; {\bf [34]} Baluci{\'n}ska-Church et al.~2004,
    MNRAS, 334, 338; {\bf [35]} Deutsch et al.~1998, ApJ, 493, 775,
    NGC6441; {\bf [36]} King et al.~1993, ApJ, 413, L117; {\bf [37]}
    Stella et al.~1987, ApJ, 312, L17; {\bf [38]} Hertz \& Wood 1985,
    ApJ, 290, 171; {\bf [39]} Stacey et al.~2012, ApJ, 751, 62; {\bf
      [40]} Engel et al.~2012, ApJ, 747, 119; {\bf [41]} Heinke et
    al.~2001, ApJ, 562, 363; {\bf [42]} Seward et al.~1976, MNRAS,
    175, 39P; {\bf [43]} Ferraro et al.~2000, ApJ, 542, L29; {\bf
      [44]} Homer et al.~1996, MNRAS, 282, L37; {\bf [45]} Bailyn et
    al.~1991, ASP Conf. Series, 13, 363; {\bf [46]} Makishima et
    al.~1981, ApJ, 247, L23; {\bf [47]} Cackett et al.~2006, MNRAS,
    369, 407; {\bf [48]} Swank et al.~1977, ApJ 212, L73; {\bf [49]}
    Revnivtsev et al.~2002, Astronomy Letters, 28, 237; {\bf [50]}
    Markwardt \& Swank 2010, IAUC 7454; {\bf [51]} Heinke et al.~2003,
    ApJ, 590, 809; {\bf [52]} Ferraro et al.~2015, ApJ, 807, L1; {\bf
      [53]} Bordas et al.~2010, ATel 2919; {\bf [54]} Pooley et
    al.~2010, ATel 2974; {\bf [55]} Riggio et al.~2012, ApJ, 754, L11;
    {\bf [56]} Papitto et al.~2011, A\&A, 526, L3; {\bf [57]} Testa et
    al.~2012, A\&A, 547, A28; {\bf [58]} Wijnands et al.~2012, ATel
    4242; {\bf [59]} Bahramian et al.~2014, ApJ, 780, 127; {\bf [60]}
    Predehl et al.~1991, A\&A, 246, L21; {\bf [61]} in 't Zand et
    al.~2003, A\&A, 406, 233.}
  \end{table}

\section{Faint sources in globular and open clusters} \label{sec_faint}

\noindent
{\em Globular clusters}\\ With the {\em Einstein Observatory} and
later {\em ROSAT}, X-ray sources in globular clusters could be studied
with better spatial resolution and sensitivity than previously
(e.g.~\cite[Hertz \& Grindlay 1983]{hertgrin83}, \cite[Verbunt
  2001]{verb01}). Despite the improved resolution, positional errors
were typically so large that identifying optical or UV counterparts
remained challenging. {\em Chandra} has brought about a radical
change. Early {\em Chandra} observations of globular clusters revealed
an overwhelming abundance and variety of faint sources; see e.g.\,the
initial work on 47\,Tuc by \cite[Grindlay et
  al.~(2001)]{grinea01}. Four main faint ($L_X \lesssim 10^{33}$ erg
s$^{-1}$) source classes can be distinguished in globular clusters: i)
quiescent neutron-star LMXBs or qLMXBs, which are often classified as
such based on their soft X-ray spectra; ii) white dwarfs accreting
from low-mass companions, also called cataclysmic variables or CVs;
iii) millisecond pulsars, sometimes isolated, often in binaries; iv)
magnetically active binaries, which are detached or contact binaries
that emit coronal X rays as a result of magnetic activity on a rapidly
spinning main-sequence or subgiant star that is tidally locked to its
companion. These active binaries or ABs are the most abundant X-ray
source class in globular clusters, and are also the X-ray faintest of
the four source types. The thermal X-ray emission of qLMXBs can be
valuable for constraining the equation of state of dense matter,
especially when the distances are well-known as is the case for
globular clusters systems (e.g.~\cite[Bogdanov et
  al.~2016]{bogdea16}).

Until recently, it was argued that most black holes that form in a
cluster are not retained for more than $\sim$1 Gyr, but are ejected
from the cluster as a result of dynamical interactions. Driven by both
theoretical and observational work, this view has now changed (for a
review see \cite[Kremer et al.~(2019)]{kremea19}, this volume). There
is no candidate stellar-mass black hole among the luminous LMXBs in
globular clusters (possibly because of the small overall sample). But
in the past years, several candidate stellar-mass black holes have
been identified both among the lower-luminosity X-ray sources (see
e.g.~\cite[Shishkovsky et al.~2018]{shisea18}) and through optical
surveys (the candidate found by \cite[Giesers et al.~(2018)]{giesea18}
does not match with any known X-ray source). The system X\,9 in
47\,Tuc (\cite[Miller-Jones et al.~2015]{millea15}), previously
classified as a bright CV, is particularly interesting, not only
because of the possible black-hole nature of the accretor, but also
because of its compact donor. Periodicity in the X-ray light curve of
$\sim$28 min shows it is an ultra-compact binary, which only leaves
room for a degenerate donor (\cite[Bahramian et
  al.~2017]{bahrea17}). The optical spectrum lacks any sign of
hydrogen (\cite[Tudor et al.~2018]{tudoea18}). It has been suggested
that X\,9 could have formed by the tidal capture of a horizontal
branch star by black hole. In this scenario, the expulsion of the
giant's envelope during the subsequent common-envelope phase, left a
white dwarf, which has now become the mass donor (\cite[Church et
  al.~2017]{churea17}).

\newpage
\noindent
{\em Old open clusters}\\ Before {\em Chandra}, several nearby
($\lesssim1$ kpc) old ($\lesssim4$ Gyr) open clusters had already been
studied with {\em ROSAT} (e.g. \cite[Belloni et al.~1998]{bellea98},
\cite[Belloni \& Tagliaferri 1998]{belltagl98}), but {\em Chandra} and
also {\em XMM-Newton} have brought the more distant---and often
older---open clusters within reach (e.g. \cite[Gosnell et
  al.~2012]{gosnea12}, \cite[van den Berg et al.~2013]{vdbergea13},
\cite[Vats \& van den Berg 2017]{vatsvand17}, \cite[Vats et
  al.~2018]{vatsea18}). Not only does this provide an opportunity to
study more comprehensively if, and how, the X-ray--source population
in open clusters evolves in time; it also narrows the age gap that
exists in X-ray studies of old open and globular clusters, and offers
a view how e.g.~cluster structure, stellar density and metallicity may
affect an old cluster's X-ray emission. Most X-ray sources in old open
clusters are ABs, and some CVs have been detected as well.  It is not
surprising that no confirmed LMXBs (neither luminous nor quiescent)
have been found in old open clusters, where present-day stellar
densities are too low to create them dynamically. Although the number
density of LMXBs in the Galactic field is uncertain, the total mass
represented by the Galactic old open clusters is very likely too low
to expect any primordial LMXBs. A group of cluster X-ray sources that
are poorly understood are the sub-subgiants, and the blue and yellow
stragglers. These are found in regions of the color-magnitude diagram
that are not occupied by single or binary cluster stars that have
followed ``standard'' evolution paths. Their X-ray properties resemble
those of ABs, and while their formation and present evolutionary state
may be unclear, the X-rays may be the result of rotation-enhanced
magnetic activity---e.g.~due to tidal interaction, or perhaps spin-up
after an episode of accretion or a merger event. The latter scenario
was suggested to explain the X-rays from the yellow straggler S\,1237
in M\,67, which is likely composed of an evolved blue straggler and a
turn-off star (or another blue straggler) in a $\sim$698 day orbit
(\cite[Leiner et al.~2016]{leinea16}). It is interesting to point out
that X-ray sources like S\,1237 have not been found in globular
clusters (yet). This may be the result of the higher stellar densities
in globular clusters, which could cause binaries in orbits as wide as
that of S\,1237 to be disrupted.

\vspace{0.3cm}
\noindent
{\em Source classification}\\ Our current understanding of the faint
X-ray sources (or, more generally, of the close binaries) in old
clusters can partly be attributed to detailed complementary work at
other wavelengths. For example, optical and/or (near-)ultraviolet
(NUV) work with {\em HST} has been indispensable to identify, classify
and characterize {\em Chandra} sources (e.g.~\cite[Edmonds et
  al.~2003a]{edmoea03a}, \cite[Edmonds et al.~2003b]{edmoea03b},
\cite[Cadelano et al.~2017]{cadeea17}, \cite[Hare et
  al.~2018]{hareea18}, \cite[Zhao et al.\ 2019]{zhaoea19}). Even with
     {\em Chandra}, the source error circles in globular clusters
     often contain more than one candidate counterpart.  Looking for
     blue, $H\alpha$-bright or variable {\em HST} counterparts inside
     the small {\em Chandra} error circles has proven to be a powerful
     way to identify the true counterparts. Accreting binaries usually
     stand out in their optical/NUV colors that are typically much
     bluer, or brighter in $H\alpha$, due to the contribution from an
     accretion disk, or (in CVs) from the white dwarf itself. ABs as
     well show $H\alpha$-excess emission from an active
     chromosphere. For open clusters, the WOCS survey
     (\cite[Mathieu~2000]{math00}) has given valuable information on
     photometry, radial velocities, and proper motions for a select
     number of clusters. With the help of {\em Gaia} proper motions,
     membership can now be established for X-ray sources in a much
     larger number of open clusters, provided the optical counterparts
     can be confidently identified.

Also without a multi-wavelength approach we can get some insight into
the nature of the X-ray sources, even if they are too faint to
constrain their X-ray spectral properties. At a distance of $\sim$1.8
kpc, M\,4 is the globular cluster nearest to us. The {\em Chandra}
observations of M\,4 are the deepest X-ray images of a globular
cluster so far. With a combined sensitivity of several times $10^{28}$
erg s$^{-1}$, these observations are especially suited to study
ABs. Only the observations of 47\,Tuc and NGC\,6397 come close to this
limit (but they do not reach it); X-ray studies of other clusters do
not go deeper than, typically, $10^{30-31}$ erg s$^{-1}$. \cite[Pooley
  (2016)]{pool16} gives a preview of the M\,4 results, showing (in
their Fig.~2) that the radial distribution of X-ray sources within the
half-mass radius is shallower than those in 47\,Tuc and
NGC\,6397. Further analysis shows that this can largely be attributed
to sources fainter than $\sim$$10^{29}$ erg s$^{-1}$, which lie below
the 47\,Tuc and NGC\,6397 detection limit.  This implies that those
faint sources are overall less massive, if the cluster relaxation time
scale is short enough for mass segregation to take effect. This is as
expected if those fainter sources are dominated by ABs, as opposed to
CVs, MSPs or qLMXBs, which are typically heavier because they contain
a compact object.

\vspace{0.3cm}
\noindent
{\em Effects of dynamical encounters}\\ With a larger sample of faint
sources, it became possible to investigate whether the correlation
between source numbers and encounter rate also extends to the qLMXBs
and other faint source classes in globular clusters. A correlation was
indeed found for the qLMXBs (e.g.~\cite[Heinke et al.~2003]{heinea03},
\cite[Pooley et al.~2003]{poolea03}). \cite[Pooley \& Hut
  (2006)]{poolhut06} were the first to show observationally that
dynamics likely also plays a role in the formation of CVs. On the
other hand, various authors (e.g.~\cite[Bassa et al.~2008]{bassea08},
\cite[Lu et al.~2009]{luea09}) used {\em Chandra} observations to show
that globular clusters contain primordial binaries as well. The level
of dynamical activity is expected to be much lower in old open
clusters, but the effect of encounters cannot be ignored altogether
with compelling evidence coming from blue stragglers. Detailed,
long-term observations indicate that the majority of blue stragglers
in NGC\,188, M\,67, and NGC\,6819 formed as a result of mass transfer
in close binaries (see \cite[Geller~(2016)]{gell16} for a summary and
references). However, there are also several blue stragglers
(including some faint X-ray sources) for which encounters probably
played a role in their formation (e.g.~\cite[van den Berg et
  al.~2001]{vdbergea01}, \cite[Mathieu \& Geller 2009]{mathgell09}).

More detailed insights into how dynamics may affect CVs in globular
custers has come from a comparison of {\em Chandra}-selected CVs in
the core-collapsed clusters NGC\,6397 and NGC\,6752 (\cite[Cohn et
  al.~2010]{cohnea10}, \cite[Lugger et al.~2017]{luggea17}), with
those in the non-collapsed cluster 47\,Tuc (\cite[Rivera Sandoval et
  al.~2018]{riveea18}). The CVs in 47\,Tuc show a single-peaked
distribution of their absolute optical magnitudes, whereas the CVs in
the core-collapsed clusters show bimodal distributions. In the latter,
the bright CVs are also found to be more massive and more concentrated
towards the cluster center than the faint ones. It was suggested that
the bright CVs are still young, and were dynamically formed near the
center of the core-collapsed clusters where interactions are assumed
to occur frequently. The fainter systems, on the other hand, are old
(and therefore less luminous as the donor masses dwindle and the
accretion rates drop), and less concentrated as a result of one or
more scattering encounters that pushed them further out. But recent
numerical simulations have provided a different
perspective. \cite[Belloni et al.~(2019)]{bellea19} find that
encounters only play a limited role in the formation of CVs, and that
most CVs that are currently observed, evolved from primordial
binaries. According to their results, the bimodal magnitude
distribution is mainly the natural result of CV evolution driven by
angular momentum loss, where the young systems are still more bright
and massive and have longer periods, and the older systems have
already evolved to short periods and have become fainter and less
massive. Whether these bright CVs show a different radial distribution
than the faint ones is determined by the relaxation time of the host
globular clusters. If it is short (as is the case in core-collapsed
clusters), the bright systems will naturally show a different, more
concentrated distribution than the faint ones, since mass segregation
has had time to drive them to different locations in the cluster. The
fact that the contribution of primordial CVs to the total observable
CV population exceeds the contribution from dynamically formed CVs,
then implies that the correlation between the number of CVs and the
normalized (to the cluster mass) encounter rate should be weak or
absent. This appears to be at odds with the findings by \cite[Pooley
  \& Hut (2006)]{poolhut06}. Another outcome of these simulations is
that the destruction of CV progenitors, which are wide main-sequence
binaries, outweighs the dynamical production of new CVs.

Do we see observational proof for binary destruction or disruption?  A
number of results may indeed point in that direction. By
parameterizing the probability that a binary experiences an encounter
{\em after} it has been dynamically formed ($\gamma$), \cite[Verbunt
  \& Freire (2014)]{verbfrei14} showed that $\gamma$ reasonably
describes which clusters are more likely to host radio pulsars that
appear to have been affected by secondary encounters (e.g~isolated or
slow pulsars, or products of exchange encounters). The analysis of
globular-cluster {\em Fermi} data by \cite[De Menezes et
  al.~(2019)]{demeea19} suggests a similar correlation, in the sense
that the inferred number of MSPs goes down for clusters with higher
$\gamma$ values (suggesting that potential progenitor systems are
disrupted before an MSP is formed).  We can also consider the issue of
binary destruction in terms of the overall number of X-ray sources per
unit mass. As has been shown by e.g.~\cite[Bassa et
  al.~(2008)]{bassea08}, the number of X-ray sources scales with both
the cluster mass (reflecting the primordial binaries) and the rate of
encounters (reflecting dynamically formed sytems). Fig.~9 in \cite[Lan
  et al.~(2010)]{lanea10} is a graphical representation of this
result. Going down to the regime of small cluster mass and few
encounters where clusters like E\,3 or NGC\,6535 are found, no or very
few X-ray sources are observed and expected (\cite[Lan et
  al.~2010]{lanea10}, Kong et al. private communication). The
present-day stellar densities in old open clusters are much lower
still. Extrapolating from the globular clusters, and applying the same
X-ray luminosity limit, we do not expect to see any X-ray sources in
old open clusters. But in fact, many ABs and some CVs that are solid
cluster members have been identified in old open clusters. Therefore,
old open clusters have more X-ray sources than the lowest-density
globular clusters, which are similar in mass or only several times
more massive (and, as outlined in Section~\ref{sec_xem}, this
difference also persists for higher-density globular clusters). Since
the current rate of encounters in these sparse globular clusters is
very low, it is not clear whether binary destruction is the (only)
cause for this discrepancy. On the other hand, it is important to keep
in mind that the encounter rate that is estimated from the current
cluster parameters may be very different from the rate in the past,
when the cluster core may have been much denser.

\section{X-ray emissivity} \label{sec_xem}

We can look at the X-ray emission of old star clusters in a larger
context and, instead of studying sources individually, compare the
combined X-ray emission from old stellar populations in general. The
combined X-ray emission per unit of mass, or X-ray emissivity, has
been measured for other environments besides globular and old open
clusters, such as various other parts of the Galaxy (the local
neighborhood, parts of the bulge, and the Galactic center
region--e.g.~\cite[Sazonov et al.~2006]{sazoea06}, \cite[Muno et
  al.~2006]{munoea06}, \cite[Revnivtsev et al.~2007a]{revnea07b},
\cite[Hong et al.~2009]{hongea09}), and extragalactic populations
(e.g.~\cite[Revnivtsev et al.~2007b]{revnea07a}, \cite[Ge et
  al.~2015]{geea15}). However, due to the different analysis methods
and adopted energy bands, it is often not straightforward to compare
the results of these studies.

Using {\em ROSAT} observations, \cite[Verbunt (2001)]{verb01} showed
for a large sample of globular clusters that they are less X-ray
luminous per unit mass than the open cluster M\,67 (when luminous
X-ray binaries are disregarded), possibly/partly due to binary
destruction, differences in metallicity or age. Also, as a result of
the shorter relaxation times of open clusters, and of their orbits in
or through the Galactic plane, open clusters lose stars at a faster
rate than globular clusters. This could artificially boost the X-ray
emission per unit of mass that we see today if the binaries that
evolve to become X-ray sources are preferentially kept. Of all old
open clusters studied at the time, the X-ray source population of
M\,67 was best characterized.  Based on deeper {\em Chandra}
observations, and taking advantage of better source classifications,
\cite[van den Berg et al.~(2013)]{vdbergea13} looked at this in more
detail and compared the number of CVs and ABs (with
$L_X\gtrsim10^{30}$ erg s$^{-1}$ and inside the half-mass radius) per
unit of mass in the old open clusters M\,67, NGC\,6819, and NGC\,6791
and the globular clusters 47\,Tuc and NGC\,6397. They found that both
CVs and ABs are underabundant in globular clusters. On the other hand,
they also showed that M\,67 may be an outlier among the old open
clusters, having relatively more ABs than the other open
clusters. This underlined the need to study the X-ray source
populations in more old open clusters to be able to put any
conclusions on firmer grounds.

\cite[Ge et al.~(2015)]{geea15} investigated the X-ray emissivities of
a somewhat larger, but still quite limited, number of old open and
globular clusters but extended the comparison to environments where
dynamical interactions are presumably irrelevant. Excluding old
extragalactic populations where hot gas or recent star formation
contribute to the X-ray emission, they find that the X-ray
emissivities of dwarf galaxies, the bulge of M\,31 and the local
neighborhood are consistent. From this they conclude there may be a
quasi-universal value of the X-ray emissivity for old, dynamically
inactive stellar populations. Although their results show that the
X-ray emission of most globular clusters in their small sample
(47\,Tuc, NGC\,6397, NGC\,6266) also agree with this value, they
acknowledge that more globular clusters need to be studied in order to
understand whether and how cluster dynamics could boost or suppress
the X-ray emission. The open clusters M\,67 and NGC\,6791 were again
found to be more X-ray luminous than the other environments with both
higher {\em and} lower stellar densities. This implies that the
observed differences in X-ray emissivity cannot be attributed to the
dynamical destruction of binaries (alone). The other outlier in the Ge
et al.~sample is $\omega$\,Cen; however, this cluster may have a very
different origin than the other Galactic globular clusters.

\cite[Cheng et al.~(2018)]{chen18} compare the X-ray emissivity of a
large sample of Galactic globular clusters. Specifically, their work
looks at the correlation with encounter rate (per unit of mass), and
revisits the earlier result by e.g.~\cite[Pooley \& Hut
  (2006)]{poolhut06} that more X-ray sources are found in clusters
where one expects more interactions. According to the analysis by
Cheng et al., the trend of increasing X-ray emissivity with encounter
rate becomes much less significant, which they take as a sign that the
contribution of primordial binaries is more important than previously
thought. This finding is in line with the theoretical findings by
\cite[Belloni et al.~(2019)]{bellea19} whose numerical simulations
indicate that the primordial CVs dominate the present-day sample of
observable CVs (see last part of Section~\ref{sec_faint}). A new study
by Heinke et al.~(in preparation) also analyzes the X-ray emissivity
of a large sample of globular clusters and several old open clusters,
focusing on investigating possible correlations with metallicity, age
and binary fraction.

\vspace{0.3cm}
\noindent
In summary, studies of the collective X-ray emission of old star
clusters and detailed investigations of individual sources complement
each other to reach a better understanding of the internal cluster
dynamics, binary evolution, and the overall population of short-period
(i.e.~X-ray--emitting) binaries.

\end{document}